\documentstyle[aps,epsfig,float]{revtex}

\begin{document}
\title{Charge Conjugation and Pairing in a model Cu$_{5}$O$_{4}$ Cluster}
\author{Michele Cini, Adalberto Balzarotti, and Gianluca Stefanucci}
\address{Istituto Nazionale di Fisica della Materia, Dipartimento di Fisica,\\
Universita' di Roma Tor Vergata, Via della Ricerca Scientifica, 1-00133\\
Roma, Italy}
\maketitle

\begin{abstract}
Highly-symmetric three-band Hubbard Cu-O clusters have peculiar properties
when the hole number is such that they admit W=0 hole pairs. These 
are two-hole eigenstates of the on-site Hubbard repulsion with eigenvalue 0, get bound by
correlation effects when dressed by the interaction with the background, and
cause superconducting flux quantization. We study the Cu$_{5}$O$_{4}$
cluster by exact diagonalization and show that bound electron pairs of $^{1}$%
B$_{2}$ symmetry are obtained at an appropriate filling, and quantize flux
like the hole pairs. The basic mechanism for pairing in this model is the
second-order exchange diagram, and an approximate charge conjugation
symmetry holds between electron and hole pairs. Further, the flux
quantization property requires that the W=0 pairs of $d$ symmetry have $s$
symmetry couterparts, still with W=0; the former are due to a spin
fluctuation, while the latter arise from a charge fluctuation mechanism. The
simultaneous existence of both is an essential property of our model and is
required for any model of superconducting $d$ pairs.
\end{abstract}


\noindent

\narrowtext
\twocolumn
\section{INTRODUCTION}

Increasing experimental evidence obtained in several cuprate superconductors
has convincingly demonstrated that the pairing state of these materials has $%
d-wave$ symmetry \cite{a} and that the pairs exist above the critical
temperature either in the form of superconducting fluctuations or preformed
pairs. The latter aspect is apparent in the underdoped (normal) region in
which a clear pseudogap essentially of the same magnitude as the
superconducting gap is measured\cite{b}. All these signatures put strict
constraints to any microscopic model of the cuprates. Any theory of the
paired state must predict the correct symmetry and doping dependence of the
binding energy of the pair, but the pairing mechanism must be
doping-independent and robust enough to survive superconducting fluctuations
well into the normal state, far from optimum doping.

In BCS theory, the first-order repulsion between like charges is overcome by
the second order interaction with phonons. In high-$T_{C}$ superconductors
the electron-phonon interaction is strong and phonons must be expected to
contribute in an important way to the pairing interaction, although their
task looks harder because the repulsion integral U is large (several eV).
However, the straightforward idea that the hight-$T_{c}$ phenomena are just
a rescaled version of BCS theory is not granted. The role of phonons may be
important, but is different, and some other ingredient is essential. First,
due to the planar C$_{4v}$ symmetry of these materials there is actually no
repulsion barrier to overcome. In a series of papers\cite{c,d,e} we have
introduced the two-hole $^{1}$B$_{2}$ eigenstates with zero Coulomb on-site
repulsion (the so called W=0 eigenstates). This is based on a standard
two-dimensional Hubbard description of the particle correlations in the
copper-oxide plane. The model has been numerically implemented in real space
on highly symmetric Cu-O clusters containing up to 21 atoms and there the
W=0 pairs arise from degenerate $(x,y)$ levels. We have shown that when the
first order interaction vanishes the net effect of the interaction is
attractive in the physical parameter range. We used the smallest hole number
($n$=4 holes) necessary to form the W=0 pair on the degenerate $e(x,y)$
one-hole level of the clusters. Also, we have shown that although U is not
small the dressing of W=0 pairs is perturbative; by exact diagonalization
and diagrammatic analysis we demonstrated that the model leads to pair
attraction of the order of tens of meV arising from virtual spin-flip
excitations. Thus, rather than proposing a model of pairing, we have pointed
out that the most standard description of electron correlations in the Cu-O
plane already leads to pairing in clusters when the symmmetry of the problem
is fully allowed. Remarkably, the off-site repulsive interactions, when
included, tend to enhance the effect somewhat \cite{d}, so we devote the
present study to the on-site interaction effects for simplicity. The
mechanism we are considering is only a part of the story, but it seems to be
the most peculiar part, being related to nothing but the C$_{4v}$ symmetry.
As such, the mechanism itself is doping-independent (although its effects do
depend on doping). For similar reasons here we wish to make abstraction from
phonon effects to see how far the simplified mechanism can account for
reality by itself. We believe that a mechanism which predictably gets
attraction out of repulsion is by itself of theoretical interest.
Superconductivity is a much more complex problem than pairing, and the
thermodynamic limit of the present model is currently under consideration;
however, we pointed out\cite{e} that the magnetic flux is quantized by our
pairs in the same way as it is in the type II superconductors.

In the
present paper we extend the analysis of Ref.\cite{e} by diagonalizing the Cu$%
_{5}$O$_{4}$ cluster with increasing number $n$ of holes. Eventually, we
find electron pairing, again with a binding energy $\Delta $(n) of a few
tenth of eV in the physical parameter space. Electron pairing is actually
realized \cite{f}, e.g., in the $T'$ structure of (Nd,Ce)$_{2}$CuO$_{4}$ exhibiting
superconductivity. The 
$T'$ structure of this compound is different from the $T$ structure 
of La$_{2}$CuO$_{4}$, but is still characterized by CuO$_{2}$ planes 
\cite{tp}.Our main point here is that electron pairs and hole pairs
are related by an approximate charge conjugation symmetry and the very same
basic mechanism or diagram is operating in both cases. Further, we
demonstrate how the two different symmetries (A$_{1}$ and B$_{2}$) of W=0
singlet pairs allowed by our theory produce the superconducting flux
quantization phenomenon, both for electron and hole pairs, in the respective
doping regimes, and are thus both necessary for a serious proposal of a
mechanism of pairing.

\section{ CHARGE CONJUGATION}

By a canonical transformation from holes to electrons, the three-band
Hubbard Hamiltonian we considered previously\cite{c} becomes: 
\begin{eqnarray}
H=\sum_{i}(2\varepsilon _{i}+U_{i})-\sum_{i\sigma }(\varepsilon
_{i}+U_{i})a_{i\sigma }^{+}a_{i\sigma }\nonumber\\-\sum_{<i,j>\sigma }t_{ij}a_{i\sigma
}^{+}a_{j\sigma }+\sum_{i}U_{i}n_{i+}n_{i-},  \eqnum{1}
\end{eqnarray}
with a$_{i\sigma }^{+}$=c$_{i\sigma }$ and a$_{i\sigma }$=c$_{i\sigma }^{+}$
creating or destructing an electron of spin $\sigma $= $\uparrow ,\downarrow 
$ at site $i$, respectively. Thus, in the electron representation the signs
of the site and hopping integrals are reversed. The single-electron energy
levels of the Cu$_{5}$O$_{4}$ cluster have been computed in the physical
range of parameters (in eV) used before, $i.e.$, U$_{d}$=5.3,U$_{p}$= $6$,
t=1.3,$\varepsilon _{d}$=0,$\varepsilon _{p}$=3.5 and the results are
displayed in Fig.1 as a function of the oxygen-oxygen hopping integral t$%
_{ox}$. With respect to the hole case \cite{c}, the sequence of levels is
such that bonding and antibonding states are interchanged. According to
symmetry arguments \cite{d}, singlet pair eigenstates of the Hamiltonian H
with zero eigenvalue of the interaction part and $^{1}$B$_{2}$ symmetry
exist when two particles sit on a degenerate $e(x,y)$ level. In the hole
representation the antibonding state of $e$ symmetry contains two holes when
14 holes are present in the cluster. Its charge-conjugate counterpart 
(since the dimensionality of the one-body basis is 18) 
corresponds to a total of 4 electrons in the cluster, two of them filling
the lowest a$_{1}$ bonding state and the other two sitting on the next $%
e(x,y)$ level of Fig.1.

\begin{figure}
\begin{center}
\epsfig{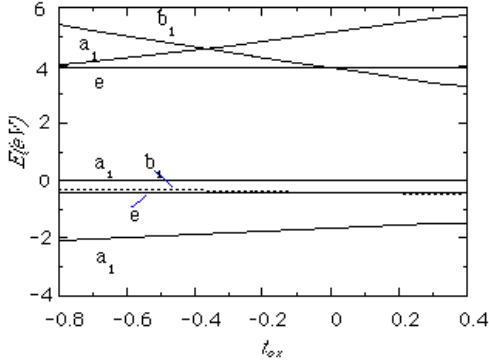}\caption{\footnotesize{
	One-electron energy levels of the Cu$_{5}$O$_{4}$ cluster versus t$%
_{ox}$. Parameters are given in the text. The levels are labelled according
to the representations of the C$_{4v}$ Group.}}
\end{center} 
\end{figure}

In the symmetric Cu$_{5}$O$_{4}$ cluster we previously found \cite{c} that
the exact diagonalization of the Hamiltonian matrix with four holes in a
wide range of negative t$_{ox}$ values for the $^{1}$B$_{2}$ singlet gives a
negative pairing energy $\Delta $(4), where $\Delta $(n) =E(n)+E(n-2)-2
E(n-1), and E(n) is the ground state energy with n holes. Therefore one
expects that $\Delta $(14) is negative while $\Delta $(12), which
corresponds to the complete filling of the upper e(x,y) level with $^{1}$A$%
_{1}$symmetry, is highly repulsive. Typical energies $\Delta $ computed with
t$_{ox}$=- 0.2 eV and different hole numbers are listed in Table 1.

The t$_{ox}$ dependence of $\Delta $ computed using an enhanced Lanczos
diagonalization routine is shown in Fig.2 and Fig.3 for $\Delta $(4) and $%
\Delta $(14) and for $\Delta $(10) and $\Delta $(12), respectively. Filling
the levels with $n$ holes, pairing occurs when the uppermost pair is a W=0
singlet. It is apparent that a considerable degree of symmetry under e-h
exchange exists in the cluster and that the pairing interaction is not
restricted to a single value of doping. The electron-hole symmetry is not
exact, because the one-electron energy level spectrum of Fig. 1 is not
invariant when up is exchanged with down and left with right 
($t_{ox}\rightarrow -t_{ox}$ and $E \rightarrow -E$); however one
could have predicted the negative $\Delta $ and its magnitude by the
approximate symmetry. In real systems electron superconductivity has been
first reported \cite{f} in the Nd$_{2}$CuO$_{4}$ doped with Ce with a
maximum T$_{c}$=24 K for optimal doping concentration x={}0.15, similarly to
more common hole superconductors. Nd$_{2-x}$Ce$_{x}$CuO$_{4}$ is an example
of high-$T_{c}$ cuprate in which charged carriers are electrons. 
\begin{figure}
\begin{center}
	\epsfig{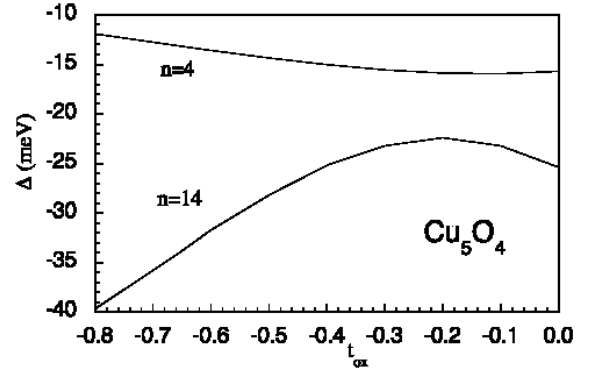}\caption{\footnotesize{
	Pairing energy $\Delta $(n)=E(n)+E(n-2)-2 E(n-1) of the Cu$_{5}$O$%
_{4} $ cluster with n=4 and n=14 holes.}}
\end{center} 
\end{figure}
\begin{figure}
\begin{center}
	\epsfig{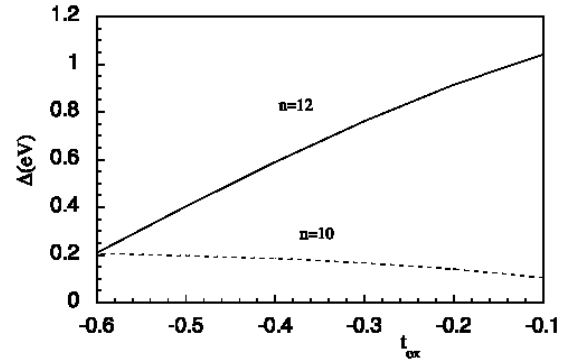}\caption{\footnotesize{
	$\Delta (n)$ as a function of t$_{ox}$ for n=10 and n=12.}}
\end{center} 
\end{figure}
The spin-flip diagram responsible for the pairing has been identified before%
\cite{e}. 
The shift of the pair energy was  found, after considerable 
algebra, in terms   of the one-body site amplitudes. When $\Delta(4)$ 
is computed, the background holes are just 2, and occupy an orbital 
of $a_{1}$ symmetry, which we denote by $a$. Empty states (antibonding 
and nonbonding) of the same 
symmetry will be denoted by $a'$, and we shall write $b$ for the empty 
states of $b_{1}$ symmetry.
To second-order one obtains for $\Delta $ the following
expression: 
\begin{equation}
\Delta =-2\left[ \sum\limits_{b}\frac{W(a,b,x,x)^{2}}{(\epsilon
_{b}-\epsilon _{a})}-\sum\limits_{a^{\prime }}\frac{W(a,a^{\prime },x,x)^{2}%
}{(\epsilon _{a^{\prime }}-\epsilon _{a})}\right]   \eqnum{2}
\end{equation}
where $W(k,l,m,n)=\langle
k_{+}l_{-}|\sum\limits_{i}U_{i}n_{i+}n_{i-}|m_{+}n_{-}\rangle $ and the sums
run only over the one-body states of $a$ and $b$ symmetry since the
contribution of the $e(x,y)$ orbitals vanishes. The final sign of $\Delta $
is determined by the relative weight of the virtual excitations to the empty
states of different point symmetry.

It is important to check the range of validity of the perturbative expansion
by comparing the results of Eq.2 with the exact diagonalization values of $%
\Delta $. We construct the normalized hole states of symmetry $a,b,e(x,y)$
whose site amplitude and energy are listed in Table II. Making use of the
above standard values of the parameters we calculate the accuracy $\delta
=2\left| \frac{\Delta _{ex}-\Delta _{appr}}{\Delta _{ex}+\Delta _{appr}}%
\right| $ for $n$=4 as a function of negative t$_{ox}$'s, i.e. in the region
of pairing. The accuracy is good (about 3\%) for small U values ( $\leq $
0.05 t ) and raises steeply to approximately 30\% for larger values of U/t
with a smooth dependence on t$_{ox}$. However, for t$_{ox}$=0 where the
pairing is strongest for the $^{1}$B$_{2}$ singlet, the accuracy is around
7\% up to U/t $\approx ${}1.

The values $U_{d}=5.3 $ eV, $U_{p}=6. $ eV  differ appreciably from 
other literature estimates\cite{mcm}, and must depend on the 
compound and doping. For La$_{2}$CuO$_{4}$, $U_{p}$=4 eV and 
$U_{d}$=10.5 eV have been recommended \cite{hyb}. None of the above results 
depends qualitatively on the precise value 
of the model parameters, since ours is basically a symmetry argument.

\section{FLUX QUANTIZATION AND PAIR SYMMETRY}

Consider the ground-state energy E($\phi $) of a two- dimensional system as
a function of the magnetic flux $\phi $ through it. This is definitely a
most compelling way of testing its superconductivity, since type II
superconductors quantize the flux for integer and half-integer multiples of $%
\phi =\frac{hc}{e}$ . If one inserts a flux tube in a loop formed by closing
the path on the external Cu of the Cu$_{5}$O$_{4}$ cluster, one finds \cite
{e} that the calculated ground state energy E(4) has a second minimum at $%
\phi _{0}$/2 well separated from a barrier from the $\phi $=0 minimum. The
point symmetry of the wavefunction changes from$^{1}$B$_{2}$($x^{2}-y^{2}$)
at $\phi $=0 to $^{1}$A$_{1}$($x^{2}+y^{2}$) at $\phi $=$\phi _{0}$/2. Owing
to the symmetry under charge conjugation and to the attractive interaction
of the E(14) state, one expects to find the specular situation in the
electron case. That this is indeed the case it is apparent from Fig.4 where
the response functions R=[E($\phi $)-E(0)]/t$_{d}$ of the hole- and
electron-doped cluster are compared. Here t$_{d}$ is an (in principle,
infinitesimal) hopping energy between the external Cu's, which is needed to
close the loop around the tube. Due to the small size of the system the
barrier preventing the condensation of pairs is finite but of comparable
height in both cases. The symmetry of the states is the same as in the hole
case, i.e., $^{1}$B$_{2}$ for $\phi $=0 and $^{1}$A$_{1}$ for $\phi $=$\phi
_{0}$/2.
\begin{figure}
\begin{center}
	\epsfig{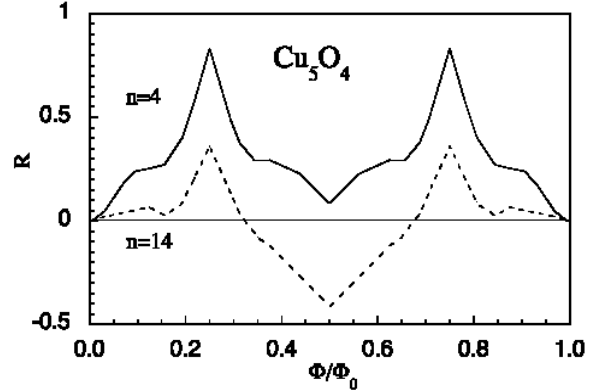}\caption{\footnotesize{
	Response function R versus the normalized flux for the Cu$_{5}$O$_{4}$
cluster with n=4 and n=14 holes.}}
\end{center} 
\end{figure}

However, the mechanism of interaction between the holes (electrons) is
basically different for $^{1}$B$_{2}$ and $^{1}$A$_{1}$ W=0 pairs. Consider
again the interesting case t$_{ox}$=0 for which all the oxygens are
equivalent and the point symmetry Group is S$_{4}$. Within the space of B$%
_{2}$ symmetry spanned by the $|$xy$>$ and $|$yx$>$ two-body states the
contribution to the scattering amplitude is zero to first-order, but finite
and negative to second order. Thus the anomalous propagator $A_{sf}\approx 
\frac{i\Delta ^{(2)}}{(\omega -2\epsilon )^{2}}$ scatters the $|$y$_{+}$x$%
_{-}>$ state out of the $|$x$_{+}$y$_{-}$ $>$ state where $\Delta ^{(2)}$ is
given in Eq.2 and $\epsilon $ is the energy of the unperturbed $e(x,y)$
states. The interaction is attractive for the lowest state of $^{1}$B$_{2}$
symmetry and involves a second order spin-flip fluctuation.

Consider now the basis of the degenerate $|$xx$>$+$|$ yy$>$ states and the $|
$bb$>$ states all having $^{1}$A$_{1}$ symmetry. The eigenvalue problem
requires the diagonalization of the $2 \times 2$ matrix: 
\begin{eqnarray}
\left| 
\begin{array}{cc}
<bb|W|bb> & <bb|W|x^{2}+y^{2}> \\ 
<bb|W|x^{2}+y^{2}> & <x^{2}+y^{2}|W|x^{2}+y^{2}>
\end{array}
\right| =\nonumber\\
\left| 
\begin{array}{cc}
\frac{U_{p}}{4} & \frac{U_{p}}{2\sqrt{2}} \\ 
\frac{U_{p}}{2\sqrt{2}} & \frac{U_{p}}{2}
\end{array}
\right|   \eqnum{3}
\end{eqnarray}
where U$_{p}$ is the on-site repulsion at the oxygen site. The lowest
eigenvalue is 0 and the W=0 pair is $\psi =-\sqrt{\frac{2}{3}}|bb\rangle +%
\sqrt{\frac{1}{3}}|x^{2}+y^{2}>$ . The upper eigenvalue is 3U/4 and the
eigenfunction $\psi =\sqrt{\frac{1}{3}}|bb\rangle +\sqrt{\frac{2}{3}}%
|x^{2}+y^{2}>$ . Thus we have a first-order mechanism involving a $charge$
fluctuation.

By looking at Fig.4 we realize that with increasing the flux $\phi $, the $%
^{1}$B$_{2}$ component of the ground state is quickly destabilized by the
vector potential, while the $^{1}$A$_{1}$ component decreases in energy and
eventually becomes energetically favored in presence of the trapped flux. A
similar change of symmetry occurs in conventional superconductors where the
Cooper wavefunction has $s$ symmetry \cite{g}. The Hubbard large U system is
stabilized by the intrinsic charge fluctuation. Since the mechanism is
robust (first order) it is conceivable that these low lying pairing states
can survive in the vortex core of the underdoped phase\cite{b}. The $^{1}$A$%
_{1}$ symmetry of the ground state in the presence of the half flux quantum
is required by general symmetry principles. Indeed, the vector potential
lowers the symmetry from C$_{4v}$ to its abelian subgroup Z$_{4}$, which
contains only the rotations. In Z$_{4}$,$^{1}$B$_{1}$ and $^{1}$B$_{2}$merge
into the same irreducibe representation and so the vector potential mixes
them, giving rise to W$\neq $ 0 pairs. This does not apply to the $^{1}$A$%
_{1}$ state, which remains single and W=0. With increasing the flux, the W$%
\neq $ 0 component of the ground state increases its energy, until it
crosses the $^{1}$A$_{1}$ pair energy. The minimum at half flux quantum
corresponds again to a $\Delta <$0 situation. It is gratifying that the
present model allows for W=0 pairs of both symmetries, because this is the
only way the superconducting flux quantization works.

\section{CONCLUSIONS}

We have investigated the symmetry properties of the W=0 pairs under charge
conjugation by performing numerical diagonalizations of the Cu$_{5}$O$_{4}$
cluster doped with a variable number of holes$\ n$ ranging from 0 to 14. We
find attractive interaction of comparable strength whenever the bonding and
antibonding degenerate levels of $x-y$ symmetry are half-filled. All the
other fillings lead to strong repulsion. The 14-hole case corresponds to the
doping with 4 electrons and thus to electron superconductivity. In all cases
the attraction is provided by the spin-flip fluctuation between $|$x$_{+}$y$%
_{-}>$ and $|$y$_{+}$x$_{-}>$ pairs. The $n$=4 and 14 hole $^{1}$B$_{2}$
states quantize the superconducting flux and the ground states in presence
of zero and one fluxon have different symmetry as in BCS superconductors.

\section{ACKNOWLEDGMENTS}

Useful discussions with A.Sagnotti are gratefully acknowledged.

\begin{center}
\bigskip 

REFERENCES
\end{center}


\bigskip

\begin{center}
TABLE CAPTIONS
\end{center}

TABLE I. $\Delta $ values of the Cu$_{5}$O$_{4}$ cluster vs the number of
holes $n$ computed with the standard set of parameters given in the text and
t$_{ox}$=-0.2 eV.

TABLE II. Symmetry, site amplitude and energy of one-hole levels for the Cu$%
_{5}$O$_{4}$ cluster. In the first row we report the Cartesian 
coordinate of each site in units of the Cu-O distance. The first 
column shows the orbitals; $a_{0}$ is the nonbonding level belonging 
to the Irrep $A_{1}$.  Minus and plus 
superscripts refer to bonding and antibonding levels, respectively. 

\bigskip

\bigskip

\begin{center}

\smallskip 

\bigskip

\bigskip TABLE I

\bigskip

\begin{tabular}{|l|l|}
\hline
$n$ holes & $\Delta $(meV) \\ \hline
4 & -15.9 \\ \hline
10 & 141.2 \\ \hline
12 & 914.7 \\ \hline
14 & -22.4 \\ \hline
\end{tabular}

\bigskip

\newpage 
TABLE II
\end{center}

\begin{tabular}{|c|c|c|c|c|c|c|c|c|c|c|c|}
\hline
 &(0,0)&(1,0)&(0,1)&(-1,0)&(0,-1)&(0,-2)&(2,0)&(0,2)&(-2,0) & energy \\ \hline
$a$ & 2$\gamma $& $\frac{\epsilon \gamma }{2t}$ & $\frac{\epsilon \gamma }{2t}
$ & $\frac{\epsilon \gamma }{2t}$ &$\frac{\epsilon \gamma }{2t}$ & $\frac{%
\gamma }{2}$ & $\frac{\gamma }{2}$& $\frac{\gamma }{2}$ & $\frac{\gamma }{2}$
 & $\epsilon _{a}^{\pm }=\frac{1}{2}\left\{ \epsilon _{p}+2t_{ox}\pm
[(\epsilon _{p}+2t_{ox})^{2}+20t^{2}]^{\frac{1}{2}}\right\} $ \\ \hline

$x$ & 0& $\frac{\alpha }{\sqrt{2}}$ &0&-$\frac{\alpha }{\sqrt{2}}$ &0&0& $%
\frac{\beta }{\sqrt{2}}$ &0&-$\frac{\beta }{\sqrt{2}}$  & $\epsilon
_{x}^{\pm }=\frac{1}{2}\left\{ \epsilon _{p}\pm [(\epsilon
_{p}^{2}+4t^{2})]^{\frac{1}{2}}\right\} $ \\ \hline
$y$ & 0&0& $\frac{\alpha }{\sqrt{2}}$ &0&-$\frac{\alpha }{\sqrt{2}}$ &- $%
\frac{\beta }{\sqrt{2}}$ &0&-$\frac{\beta }{\sqrt{2}}$ &0 & $\epsilon
_{x}^{\pm }$ \\ \hline
$b$ & 0& $\frac{\alpha _{1}}{2}$ &-$\frac{\alpha _{1}}{2}$ & $\frac{\alpha
_{1}}{2}$ &-$\frac{\alpha _{1}}{2}$ & $\frac{\beta _{1}}{2}$ &-$\frac{%
\beta _{1}}{2}$ & $\frac{\beta _{1}}{2}$ &- $\frac{\beta _{1}}{2}$  & $%
\epsilon _{b}^{\pm }=\frac{1}{2}\left\{ \epsilon _{p}-2t_{ox}\pm [(\epsilon
_{p}-2t_{ox})^{2}+4t^{2}]^{\frac{1}{2}}\right\} $ \\ \hline
$a_{0}$ &  $\frac{1}{\sqrt{5}}$ &0&0&0&0&-$\frac{1}{\sqrt{5}}$ &-$\frac{1%
}{\sqrt{5}}$ &-$\frac{1}{\sqrt{5}}$ &-$\frac{1}{\sqrt{5}}$  & 0 \\ \hline
\end{tabular}
\smallskip 

\begin{tabular}{l}
$\alpha =\frac{t}{\sqrt{t^{2}+(\epsilon _{x}^{\pm }-\epsilon _{p})^{2}}}%
;\alpha _{1}=\frac{t}{\sqrt{t^{2}+(\epsilon _{b}^{\pm }+2t_{ox}-\epsilon
_{p})^{2}}};\beta =\frac{\alpha }{t}(\epsilon _{x}^{\pm }-\epsilon
_{p});\beta _{1}=\frac{\alpha }{t}(\epsilon _{p}-\epsilon _{b}^{\pm
}-2t_{ox});$ \\ 
$\gamma =\frac{t}{\sqrt{(\epsilon _{a}^{\pm })^{2}+5t^{2}}}$%
\end{tabular}
\end{document}